\documentclass[showpacs,preprintnumbers,amsmath]{revtex4}
\usepackage{amssymb}

\usepackage{graphicx}

\begin{document}
\title{Self-localization of holes in a lightly doped Mott insulator}
\author{Su-Peng Kou$^1$ and Zheng-Yu Weng$^2$}
\address{$^1$Department of Physics, Beijing Normal University, Beijing, 100875, China%
\\
$^2$Center for Advanced Study, Tsinghua University, Beijing, 100084, China}

\begin{abstract}
We show that lightly doped holes will be self-trapped in an
antiferromagnetic spin background at low-temperatures, resulting in a
spontaneous translational symmetry breaking. The underlying Mott physics is
responsible for such novel self-localization of charge carriers. Interesting
transport and dielectric properties are found as the consequences, including
large doping-dependent thermopower and dielectric constant, low-temperature
variable-range-hopping resistivity, as well as high-temperature
strange-metal-like resistivity, which are consistent with experimental
measurements in the high-T$_c$ cuprates. Disorder and impurities only play a
minor and assistant role here.
\end{abstract}

\pacs{74.20.Mn, 74.25.Ha, 75.10.-b }
\maketitle

\section{Introduction}

There exist several types of electron localizations in condensed matter
physics. In the presence of disorder or impurities, waves can be localized
due to quantum interference, which is generally known as the Anderson
localization \cite{a}. In a two-dimensional (2D) system carriers are
expected \cite{abr} to be always localized with the resistance diverging
either logarithmically (``weak localization'') or exponentially (``strong
localization'') as $T\rightarrow 0$. A different kind of localization
involves the self-trapping of small polarons in strong electron-phonon
interacting systems. Both types of localizations here mainly concern
non-interacting or weakly correlated electrons.

In the high-$T_c$ cuprates, the undoped system is a Mott insulator, in which
the charge degree of freedom is totally frozen out by strong on-site Coulomb
repulsion. Its spins form an antiferromagnetic long range order (AFLRO) at
low temperatures, which quickly collapses upon hole-doping \cite{keimer}. At
small concentration, $\delta <0.05$, the doped holes as charge carriers
remain localized with the low-$T$ resistivity well fit \cite
{preyer,chen,kastner,ando} by those of traditional variable-range-hopping
(VRH) type, usually applicable to doped semiconductors. Only when the hole
concentration exceeds $0.05$ will the charge carriers be truly delocalized,
where the ground state becomes a d-wave superconducting state.

But lightly doped high-$T_c$ cuprates are of no conventional doped
semiconductors. Here they are doped Mott insulators with the majority of the
charge degree of freedom still remaining frozen. The strongly correlated
effect should thus play a crucial role in the charge transport. The issue
why doped holes should be always localized at small doping and how they
eventually become delocalized with increasing doping is a very important
question, which may also be relevant to understanding the microscopic
mechanism of high-temperature superconductivity in the cuprates.

Recently, a new kind of self-localization, which can be particularly
attributed to the Mott physics, has been proposed \cite{1hole,kou,kou1}, in
which the {\em pure} Coulomb interaction is responsible for the
self-localization of electrons near half-filling. Here disorder or
impurities is no longer as essential as in the Anderson localization,
although they may be helpful to induce a true translational symmetry
breaking at $T=0$. Such a self-localization of holes is caused by the
phase-string effect \cite{string0}, referring to a stringlike defect left by
the hopping of a hole, which is {\em irreparable} in the ground state of the 
$t-J$ model no matter whether there is an AFLRO or not at {\em arbitrary}
doping. Furthermore, it has been also shown \cite{1hole} that the charge
localization does not contradict to the photoemission experiments \cite
{arpes} in which the observed ``quasiparticle'' dispersion in the
single-particle spectral function can be well accounted for in terms of
the ``spinon dispersion''.

In principle, it is not surprising that due to the separation \cite{book}
of spin
and charge degrees of freedom in a doped Mott insulator, the charge carriers
will get localized at low doping by scattering with fluctuations from an
independent of degrees of freedom. In fact, in a different gauge-theory
approach to the $t-J$ model, the localization of charge carriers has been
also obtained \cite{yu}, by scattering to the gauge fluctuations.

In this paper, we study such a self-localization phenomenon based on an
effective phase-string description \cite{string0,string1} of the $t-J$
model. We shall focus on the spin ordered phase (not necessarily long-range
ordered) at very low doping and explore the unique charge self-localization
behavior. As shown previously, a doped hole in such a regime will induce 
\cite{kou,kou1} a dipole-like spin structure. We find that the kinetic
energy of the hole is severely frustrated by the phase string effect such
that the hole-dipole is self-trapped in real space. We discuss the
corresponding transport properties and show that the thermopower saturates
to a Heikes-like formula as the result. While the resistivity exhibits a
Mott-VRH-like behavior at low temperatures, the collapse of the hole-dipole
composites and the release of free ``holons'' at high-temperatures will lead
to a strange-metal-like behavior there. Furthermore, the existence of the
hole dipolar structure predicts a large doping-dependent dielectric
constant, which diverges at the deconfinement. All of these properties seem
to paint a consistent picture for the complex transport and dielectric
phenomena observed in the high-$T_c$ cuprates.

The remainder of the paper is organized as follows. In Sec. II, we discuss
the self-localization of doped holes in the spin ordered phase at low doping
based on the phase-string model. A renormalization group (RG) analysis will
be used to determine the phase diagram. In Sec. III, we discuss the
experimental implications of the self-localization, including the
thermopower, resistivity and dielectric constant, and make comparisons with
experimental measurements. Finally, the conclusions are presented in Sec. IV.

\section{Self-localization of holes in a doped Mott insulator}

We shall adopt the phase-string model as the microscopic description of how
doped holes move in an antiferromagnetic (AF) Mott insulator. This model is
obtained \cite{string1} as a low-energy effective theory based on the $t-J$
model, which can accurately describe AF correlations at half-filling and
possesses a d-wave superconducting ground state at doping concentrations
larger than $x_c$ ($\simeq 0.043$ at $T=0)$ \cite{kou}. What we will be
interested in the following is the non-superconducting phase below the
critical doping $x_c,$ where an AFLRO or a spin glass state persists at low
temperatures.

The existence of such a non-superconducting phase is the consequence that
the {\em long-range} AF correlations (not necessarily AFLRO) win in the
competition with the kinetic energy of holes at sufficiently low doping. In
the phase-string model, this phase will be characterized by the ``spinon
condensation''. In the following, we shall analyze in detail the behavior of
doped holes in this low-doping regime and demonstrate that the charge
carriers must be self-localized in the ground state, resulting a
spontaineous translational symmetry breaking.

\subsection{Phase-string model}

We start with the phase-string model $H_{{\rm string}}=H_h+H_s,$ which is
composed \cite{string1} of two terms: The charge degree of freedom as
characterized by the ``holon'' term

\begin{equation}
H_h=-t_h\sum_{\langle ij\rangle }\left( e^{iA_{ij}^s}\right) h_i^{\dagger
}h_j+H.c.  \label{hh}
\end{equation}
where $t_h\sim t$ and the ``holon'' operator, $h_i^{\dagger },$ is bosonic;
The spin degrees of freedom as described by the ``spinon'' term 
\begin{equation}
H_s=-J_s\sum_{\langle ij\rangle \sigma }\left( e^{i\sigma A_{ij}^h}\right)
b_{i\sigma }^{\dagger }b_{j-\sigma }^{\dagger }+H.c.  \label{hs}
\end{equation}
where $J_s\sim J$ and the ``spinon'' operator, $b_{i\sigma }^{\dagger },$ is
also bosonic.

Basic features of this model are as follows. At half filling, the gauge
field $A_{ij}^h$ can be set to zero in Eq.(\ref{hs}) and $H_s$ reduces to
the Schwinger-boson mean-field Hamiltonian \cite{aa}, which describes both
the long-range and short-range AF correlations fairly well. Upon doping, $%
A_{ij}^h$ is no longer trivial as it satisfies a topological constraint: $%
\sum_CA_{ij}^h=\pi \sum_{l\in \Sigma _C}n_l^h$ ($\Sigma _C$ denotes the area
enclosed by $C$) with $n_l^h$ denoting the ``holon'' number at site $l,$
which is interpreted as that each ``holon'' behaves like a $\pi $-fluxoid as
felt by the ``spinons''. Thus, $A_{ij}^h$ will play the key role of
frustrations introduced by holes that act on the spin degrees of freedom.
Similarly, the ``holons'' are also subjected to frustrations from the spin
background, through the gauge field $A_{ij}^s$ in Eq.(\ref{hh}). Here $%
A_{ij}^s$ satisfies a topological constraint: $\sum_CA_{ij}^s=\pi \sum_{l\in
\Sigma _C}\left( n_{l\uparrow }^b-n_{l\downarrow }^b\right) $ with $%
n_{l\sigma }^b$ denoting the ``spinon'' (with index $\sigma $) number at
site $l,$ which can be interpreted as that each ``spinon'' behaves like a $%
\pm \pi $-fluxoid as perceived by the ``holons''.

The spin and charge degrees of freedom are thus mutually ``entangled'' in
the phase-string model $H_{{\rm string}}$ based on two topological gauge
fields, $A_{ij}^h$ and $A_{ij}^s$. It has been shown \cite{zhou} that if the
holons (which are bosons) experience a Bose condensation at larger doping,
the spinons, which behave as vortices according to $A_{ij}^s$, must be
``confined'' at low temperatures. Such a ``spinon confining phase'' actually
corresponds to the d-wave superconducting phase in the model. Here, the
spinon confinement occurring in the {\em spin degree of freedom} is closely
related to the superconducting phase coherence in the {\em charge degree of
freedom}.

In contrast, if the spinons (which are also bosons) are Bose condensed at
low doping, free holons as ``vortices'' cannot live alone either, and must
be also ``confined'' \cite{kou,kou1}. Such a {\em spinon condensed phase} is
thus also known as ``holon confining phase'',{\em \ }which constitutes a
non-superconducting solution of the phase-string model in the lightly doped
regime. In the following, we will demonstrate how holons will get
self-trapped in real space in such a phase. The spin ordered phase as
characterized by the spinon condensation is therefore intimately related to
the charge self-localization (holon confining). Fig. 1 schematically
illustrates the above-mentioned two phases at different doping
concentrations.

\begin{figure}[tbp]
\begin{center}
\includegraphics{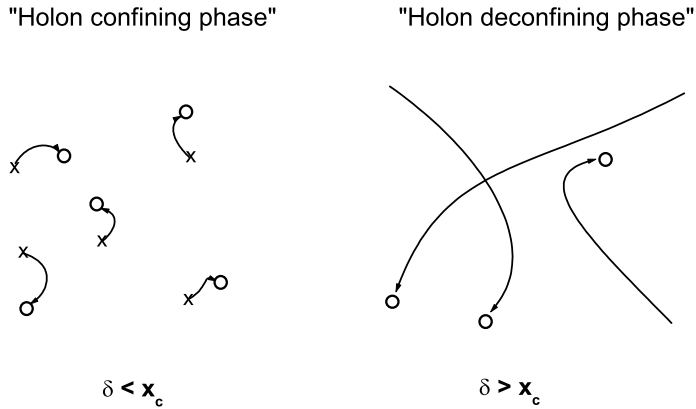} 
\end{center}
\caption{Holes are self-trapped in real space at $\delta<x_c$ by the 
phase-string effect, carrying dipolar spin configurations. Holes become
delocalized at $\delta>x_c$, with the critical doping concentration $x_c=0.043$ (see text). }
\label{fig1}
\end{figure}

\subsection{Antimerons induced by holons: ``Flux-quantization'' condition}

Let us recall that at half-filling, the Bose condensation of spinons, 
\begin{equation}
<b_{i\sigma }>\neq 0  \label{b}
\end{equation}
will naturally gives rise to an AFLRO lying in the $x$-$y$ plane \cite
{remark1} with $\left\langle S_i^{+}\right\rangle =(-1)^i<b_{i\uparrow
}><b_{i\downarrow }>$. However, once a hole is added into such an ordered
state, the energy cost associated with a {\em bare} holon, if the condensation
(\ref{b}) remains unchanged, would become {\em logarithmic ally divergent} in
terms of $H_s$\cite{kou} 
\begin{eqnarray}
\Delta E_s &\simeq &\tilde{J}_s\int d^2{\bf r}\left[ {\bf A}^h\right] ^2 
\nonumber \\
&\sim &\frac \pi 2\tilde{J}_s{\rm \ln }({L/a)\rightarrow \infty ,}
\label{des}
\end{eqnarray}
where $\tilde{J}_s\equiv J_s<b_{i\uparrow }><b_{i\downarrow }>$, ${\bf A}^h$
is the continuum version of $A_{ij}^h$, $L$ is the size of the sample and $a$
is the lattice constant.

But this is not the correct physical picture. In reality, the two-component
spinon ``superfluid'' in Eq.(\ref{b}) can {\em response} to the presence of
holons by forming a {\em spin} {\em supercurrent}. Such a spin supercurrent
can then screen out the effect of the ``magnetic fluxes'' introduced by $%
A_{ij}^h$ in $H_s,$ just like the flux-quantization phenomenon in a
superconductor, such that the resulting {\em renormalized} holon will
acquire a finite self-energy. In the following, we show how this screening
effect takes place in the phase-string model (\ref{hs}).

Define $<b_{i\sigma }>\equiv \sqrt{\rho _c^sa^2}z_{i\sigma },$ if $i\in $ $A$
sublattice, and $<b_{i\sigma }>\equiv \sqrt{\rho _c^sa^2}z_{i-\sigma }^{*},$
if $i$ $\in B$ sublattice, with $|z_{i\uparrow }|^2+|z_{i\downarrow }|^2=1.$
Here $\rho _c^s$ denotes the spinon ``superfluid'' density. Then the
condensed part of $H_s$ can be written as 
\begin{eqnarray}
H_s^{{\rm cond}} &=&-J_s\sum_{\langle ij\rangle \sigma }\left( e^{i\sigma
A_{ij}^h}\right) <b_{i\sigma }^{\dagger }><b_{j-\sigma }^{\dagger }>+c.c. 
\nonumber \\
&=&-\tilde{J}_s\sum_{i\in A,j=NN(i)}\sum_\sigma \left( e^{i\sigma
A_{ij}^h}\right) z_{i\sigma }^{*}z_{j\sigma }+c.c.  \nonumber \\
&\simeq &-\tilde{J}_s\sum_{i\in A,j=NN(i)}\sum_\sigma \left( 1+i\sigma
A_{ij}^h-\frac 12\left( A_{ij}^h\right) ^2\right) z_{i\sigma }^{*}z_{j\sigma
}+c.c.  \nonumber \\
&\simeq &E_s^0+i\tilde{J}_s\int d^2{\bf r}\sum_\sigma \sigma {\bf A}^h\cdot
\left( z_\sigma ^{*}{\bf \nabla }z_\sigma -c.c.\right) +\text{ }\tilde{J}%
_s\int d^2{\bf r}\left( {\bf A}^h\right) ^2  \label{hscond}
\end{eqnarray}
in the continuum limit, where $\tilde{J}_s=J_s\rho _c^sa^2,$ $E_s^0\equiv -4%
\tilde{J}_sN,$ and $NN$ denotes the nearest neighboring sites. The spin
supercurrent, defined by ${\bf J}^s=-\partial H_s^{{\rm cond}}/\partial {\bf %
A}^h,$ then reads

\begin{equation}
{\bf J}^s=2\tilde{J}_s\left( {\bf v}^s-{\bf A}^h\right)  \label{current}
\end{equation}
where 
\begin{equation}
{\bf v}^s\equiv -\frac i2\sum_\sigma \sigma \left( z_\sigma ^{*}{\bf \nabla }%
z_\sigma -c.c.\right) .  \label{js}
\end{equation}

Generally one expects a complicated distribution of the spin current nearby
a holon. But in a distance far away from the holon, the supercurrent ${\bf J}%
^s$ should vanish in order to ensure the finiteness of the energy cost
introduced by a holon. The same requirement has been used in a
superconductor to realize the flux quantization. By contrast, here we are
dealing with the ``flux-quantization'' in a two-component superfluid problem
with an internal gauge freedom.

By requiring the spin supercurrent ${\bf J}^s$ vanish at the boundary $%
C_\infty $ such that 
\begin{equation}
\oint_{C_\infty }d{\bf r}\cdot {\bf J}^s=0,  \label{fq0}
\end{equation}
one arrives at

\begin{equation}
\oint_{C_\infty }d{\bf r}\cdot {\bf v}^s-\oint_{C_\infty }d{\bf r}\cdot {\bf %
A}^h=0.  \label{fq1}
\end{equation}
Note that $\oint_{C_\infty }d{\bf r}\cdot {\bf v}^s=\int d^2{\bf r}\left( 
{\bf \nabla \times v}^s\right) \cdot {\bf \hat{z}}$. By introducing a unit
vector 
\[
{\bf n=}\bar{z}{\bf \hat{\sigma}}z
\]
where ${\bf \hat{\sigma}}$ is the Pauli matrix and $\bar{z}\equiv
(z_{\uparrow }^{*},z_{\downarrow }),$ one can straightforwardly show that 
\begin{equation}
\left( {\bf \nabla \times v}^s\right) \cdot {\bf \hat{z}}=\frac 12{\bf %
n\cdot \partial }_x{\bf n\times \partial }_y{\bf n.}  \label{fq3}
\end{equation}
Thus, Eq.(\ref{fq0}) finally reduces to the following ``flux-quantization''
condition 
\begin{eqnarray}
Q &\equiv &\int d^2{\bf r}\text{ }\frac 1{4\pi }\left( {\bf n\cdot \partial }%
_x{\bf n\times \partial }_y{\bf n}\right) {\bf =}\frac 1{2\pi }%
\oint_{C_\infty }d{\bf r}\cdot {\bf A}^h  \nonumber \\
&=&\frac 12N^h  \label{fq4}
\end{eqnarray}
where $N^h$ is the total number of doped holes.

Therefore, in the spinon condensed state, a holon will always induce a
``screening'' response from the spinon condensate, which is of topological
nature satisfying the ``flux-quantization'' condition (\ref{fq4}), meaning
that each holon will ``nucleate'' a spin ``meron'' configuration with a
topological charge $Q=1/2.$ Such a meron configuration may be pictured as a
spin vortex with the unit vector ${\bf n}$ lying in a spin $x$-$y$ plane at
a distance far away from the core, while, near the core, the unit vector $%
{\bf n}$ starts to tilt away from the $x$-$y$ plane and points towards the $z$%
-axis at the core center, which covers one half of the unit sphere spanned
by ${\bf n}$ once, in contrast to a skyrmion which covers the whole unit
sphere exactly once with $Q=1.$

Finally we note that such {\em holon-induced} merons are called {\em %
antimerons} in the earlier approach \cite{kou,kou1} because a holon itself
also carries a meron (vortex) in the {\em original spin space}. To see this,
let us recall that in the phase-string model, the spin operators are
expressed in terms of spinon operators by \cite{string0} 
\begin{equation}
S_i^{+}=(-1)^ib_{i\uparrow }^{\dagger }b_{i\downarrow }\exp \left[ i\Phi
_i^h\right]  \label{s+}
\end{equation}
and $S_i^{-}=\left( S_i^{+}\right) ^{\dagger },$ $S_i^z=\sum_\sigma \sigma
b_{i\sigma }^{\dagger }b_{i\sigma }.$ In the spinon condensed phase, one has 
\begin{equation}
\left\langle S_i^{+}\right\rangle =(-1)^i<b_{i\uparrow }><b_{i\downarrow
}>\exp \left[ i\Phi _i^h\right] ,  \label{s+1}
\end{equation}
which is twisted away from an AFLRO lying in the $x$-$y$ plane by the
vortices centered at holons as determined by $\Phi _i^h$. Here $\Phi _i^h$
is defined by 
\begin{equation}
\Phi _i^h=\sum_{l\neq i}\mathop{\rm Im}\ln (z_i-z_l)n_l^h~.  \label{phih}
\end{equation}
But Eq.(\ref{s+1}) describes an {\em bare} holon effect which would result
in a divergent self-energy of the holon as shown in Eq.(\ref{des}). In order
to compensate such a vortex configuration associated with a bare holon, the
condensed spinon fields have to be twisted into

\begin{equation}
<b_{i\sigma }>\rightarrow <b_{i\sigma }>\exp \left[ i\frac \sigma 2\vartheta
_i\right]  \label{bmeron}
\end{equation}
with an {\em antimeron} configuration $\vartheta _i$\cite{kou,kou1}, which
is characterized just by $z_{i\sigma }$ in the present approach, satisfying
the ``flux-quantization'' condition (\ref{fq4}). Therefore, in the
phase-string model, a renormalized holon is a composite with a bare holon
bound to an induced antimeron as demonstrated above, while in the {\em %
original spin space}, it is an object composed of a meron (holon) and an
antimeron, forming a {\em dipole }\cite{kou,kou1}. Two descriptions are
equivalent.

\subsection{Self-localization of holes: RG analysis}

According to the above analysis, the infinite self-energy of a bare holon in
the spinon condensed phase can be removed by ``nucleating'' a topological
spin antimeron configuration. This induced antimeron is an infinite-size
semiclassical object which cannot propagate based on $H_s.$ Thus, holons
will be self-trapped around the cores of these antimerons. Due to the
translational symmetry, these induced antimerons can be located anywhere in
space and therefore will result in a {\em spontaneous translational symmetry
breaking} in the spinon condensed phase.

Physically, such a self-localization of charge carriers in the low-doping
regime can be attributed to the irreparable phase-string effect created by
the motion of holes, as discussed in Ref.\cite{kou1}. The phase-string model
provides a mathematical framework to conveniently handle this effect. In
this description, the locations of the antimeron and the holon inside a
dipole constitute the starting and ending points of the motion of a holon,
and the phase-string connecting such two points has relaxed into a dipolar
picture, with a remaining branch-cut connecting two poles \cite{kou1}.

If one tries to move away the bare holon from the core of the induced
antimeron, the uncompensated spin supercurrent ${\bf J}^s$ will increase the
self-energy, representing an attractive potential which binds the bare holon
to the antimeron. According to Refs.\cite{kou,kou1}, such a potential can be
estimated by 
\begin{equation}
V\simeq q^2\ln \frac{\left| {\bf r}\right| }a  \label{vdipole}
\end{equation}
at $|{\bf r|>}a$, where $q^2=\pi \tilde{J}_s$, and ${\bf r}$ is the distance
between the holon and antimeron. At $|{\bf r|\rightarrow \infty }$, $V$
diverges in consistency with Eq.(\ref{des}).

When there are many holon-antimeron dipoles, one expects to see a {\em %
screening} {\em effect} on the confining potential by reducing $V$ to $V_{%
{\rm eff}}=\frac 1\kappa V,$ where $\kappa $ denotes the dielectric
constant. Previously it has been shown \cite{kou} that with the increase of
doping concentration, eventually a transition at $T=0$ takes place, as $%
\kappa \rightarrow \infty $ or $V_{{\rm eff}}\rightarrow 0$, from the holon
self-localization (confining) phase to a delocalization (deconfining) phase
at $\delta =x_c\simeq 0.043$ (see Fig. 1). On the other hand, with the
increase of temperature, neutral vortex-antivortex pairs like those in the $%
XY$ model can also be thermally excited, leading to a conventional
contribution to the screening effect. To distinguish such two kinds of
vortex-antivortex pairs, we shall call a dipole associated with a holon as a
charged pair and the other kind as a neutral pair. In the following, we
shall treat the screening effect due to both charged and neutral dipoles on
an equal footing based on an RG treatment \cite{kou,timm}. In contrast to
the conventional Kosterlitz-Thouless (KT) theory \cite{BKT}, here the number
of vortex-antivortex pairs remains finite even at $T=0,$ where it is equal
to $\delta N$ ({\em i.e.,} the number of holons).

The probability for the neutral dipoles with two poles separated by a
distance $r$ is controlled by the neutral pair fugacity $y_n^2(r)$. In the
conventional KT theory \cite{BKT}, the initial $y_n^2(a)=e^{-\beta E_c^n}$,
(where $\beta =\frac 1{k_BT}$ and the core energy $E_c^n\sim q^2$). But
since the charged dipole number will be always fixed at $\delta $ per site,
the probability for the charged dipoles is no longer governed by the
fugacity $y_h^2(a)=e^{-\beta E_c^h}$, (where $E_c^h$ denotes its core
energy). Instead, the initial fugacity must be adjusted accordingly and $%
E_c^h$ will turn out to be no longer so important \cite{kou,timm}.

In an RG procedure, the contributions from the dipoles with the sizes
between $r$ and $r+dr$ will be integrated out, starting from{\bf \ }$r=a$.
The renormalization effect is then represented by three renormalized
quantities, $X(r)\equiv \frac{2\pi \kappa }{\beta q^2},$ $y_n^2(r),$ and $%
y_h^2(r)$, which satisfy the following famous recursion relations \cite
{BKT,cha} 
\begin{eqnarray}
dy_h/dl &=&(2-\frac \pi X)\,y_h,  \label{rc1} \\
dy_n/dl &=&(2-\frac \pi X)\,y_n,  \label{rc2} \\
dX/dl &=&4\pi ^3(y_n^2+y_h^2)\text{,}  \label{rc30}
\end{eqnarray}
where $r=ae^l$.

Define $Y^2(l)=y_h^2(l)+y_n^2(l),$ with $Y_0^2=y_h^2(l=0)+y_n^2(l=0).$ From
Eqs.(\ref{rc1})-(\ref{rc30}), we find 
\begin{equation}
Y^2=Y_0^2+\frac 1{\pi ^3}(X-X_0)-\frac 1{2\pi ^2}\ln \frac X{X_0}
\end{equation}
where $X_0\equiv X(l=0)=\frac{2\pi }{\beta q^2}$ (with $\kappa (l=0)=1$).
The RG flow is then obtained from Eq.(\ref{rc30}) by 
\begin{equation}
l=\int_{X_0}^X\frac{dX^{\prime }}{4\pi ^3Y_0^2+4(X^{\prime }-X_0)-2\pi \ln
(X^{\prime }/X_0)}.  \label{l}
\end{equation}
The neutral pair fugacity can be determined by $y_n^2(l)=e^{-2\int_0^l(2-%
\frac \pi X)dl^{\prime }},$ which will show a similar behavior as in the
conventional KT theory.

What makes the present approach different from the conventional KT theory is
the presence of a finite density of the charged dipoles (holon-antimeron
pairs). Here, by noting $\frac{y_h^2(r)}{r^4}d^2{\bf r}$ as the areal
density of charged pairs of sizes between $r$ and $r+dr$ \cite{BKT}, we have
the following constraint 
\begin{equation}
\delta /a^2=\int_a^\infty dr\,2\pi r\,\frac{y_h^2(r)}{r^4},
\end{equation}
or 
\begin{eqnarray}
\delta /a^2 &=&\int_a^\infty dr\,2\pi r\,\frac{Y^2(r)}{r^4}-\frac 1{2\pi
^2a^2}\int_0^\infty dle^{-2l}\,y_n^2  \nonumber \\
&=&\frac 1{2\pi ^2a^2}\int_0^\infty dle^{-2l}\frac{dX}{dl}-\frac 1{2\pi ^2a^2%
}\int_0^\infty dle^{-2l}\,y_n^2.\text{ }
\end{eqnarray}
After an integral transformation, such a constraint can be rewritten as 
\begin{equation}
2\pi ^2\delta +\frac{2\pi k_BT}{q^2}+\frac 1{2\pi ^2a^2}\int_0^\infty
dle^{-2l}\,y_n^2=2\int_0^\infty e^{-2l}\,X\text{ }dl\text{.}  \label{rc3}
\end{equation}
For $\beta q^2\gg 1,$ $y_n^2<y_n^2(0)\sim \exp (-\beta q^2)\rightarrow 0,$
the probability for neutral pairs remains a small number and can be
neglected.

\begin{figure}[tbp]
\begin{center}
\includegraphics{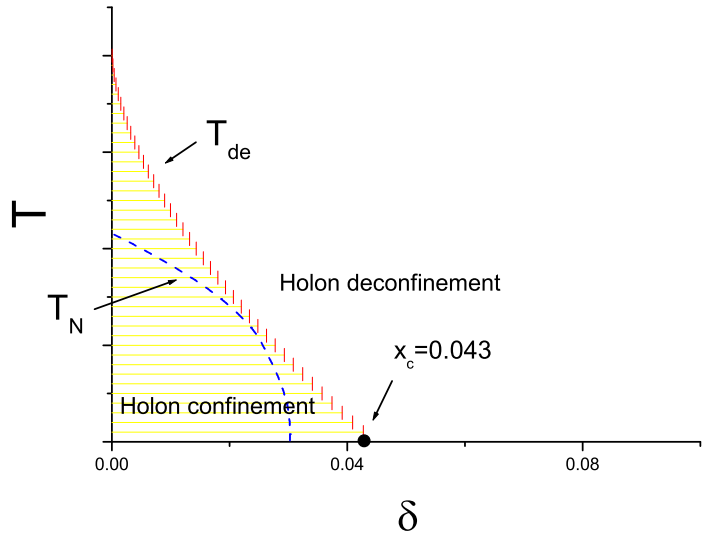} 
\end{center}
\caption{The low-doping phase diagram determined by the RG analysis. The 
boundary set by $T_{{\rm 
de}}(\delta)$ separates the 
holon confining and deconfining phases. The N\'{e}el temperature $T_{{\rm 
N}}$ is determined by introducing an interlayer coupling.}
\label{fig2}
\end{figure}

The RG flow diagram of Eqs.(\ref{rc1})-(\ref{rc30}) is as follows: The two
basins of attraction are separated by the initial values which flow to $%
X^{*}\rightarrow \frac \pi 2$ and $y_h^{*}\rightarrow 0$, $%
y_n^{*}\rightarrow 0$ in the limit $l\to \infty $. In terms of Eq.(\ref{l}),
the separatrix of the RG flows is given by 
\begin{equation}
l=\int_{X_0}^X\frac{dX^{\prime }}{4(X^{\prime }-\frac \pi 2)-2\pi \ln
(2X^{\prime }/\pi )}.  \label{rc4}
\end{equation}

``Deconfining'' temperature $T_{{\rm de}}.$ ---Based on the RG equations of (%
\ref{rc3}) and (\ref{rc4}), one can determine the critical hole density $%
\delta _{{\rm de}}=\delta _{{\rm de}}(T)$ or temperature $T_{{\rm de}}=T_{%
{\rm de}}(\delta )$ at which the charged dipoles collapse and holons are
``deconfined'' from the bound state with the antimerons and become
delocalized.

We first consider the case at $T=0.$ Since neutral vortices do not exist at $%
T\to 0$, Eq. (\ref{rc3}) reduces to 
\begin{equation}
2\pi ^2\delta _{{\rm de}}=2\int_0^\infty e^{-2l}\,X\text{ }dl=\int_0^{\pi
/2}e^{-2l}\,dX\text{ .}
\end{equation}
The critical doping $\delta _{{\rm de}}$ at $T=0$ is then numerically
determined by $\delta _{{\rm de}}(0)\simeq 0.043,$ which was previously
obtained in Ref.\cite{kou} as denoted by $x_c$. Now consider the limit at $%
\delta \rightarrow 0$, where we approximately have $X(l)\simeq X_0\simeq
X(l\rightarrow \infty )=\frac \pi 2,$ and Eq. (\ref{rc3}) becomes 
\begin{equation}
2\pi ^2\delta +\frac{2\pi k_BT_{{\rm de}}}{q^2}\simeq \frac \pi 2,
\end{equation}
which gives rise to $T_{{\rm de}}(\delta =0)\simeq \frac{q^2}{4k_B}$ and $T_{%
{\rm de}}(\delta )\simeq T_{{\rm de}}(0)-\delta \pi q^2/k_B$ at $\delta \ll
1.$

The holon ``confining'' and ``deconfining'' phases are separated by $T_{{\rm %
de}}(\delta )$ in the $T-\delta $ phase diagram as shown in Fig. 2. In the
``confining'' phase, holons are self-trapped by binding to the induced
antimerons. In the original spin space, each renormalized holon can be
regarded as a dipolar object. In this regime, AFLRO or short-ranged AF
ordering (spin glass) can still form, if the interlayer coupling is
introduced: The N\'{e}el transition temperature is obtained as $T_N(\delta
)\approx -\frac{\pi J}{k_B\ln \alpha }-3\delta J/k_B$, with $\alpha \sim
10^{-5}$ representing the ratio of the interlayer coupling $J_{\perp }/J$%
\cite{kou}, which is schematically illustrated in Fig. 2 by the dashed
curve. Previously we have also shown that a superconducting phase will set
in beyond $\delta >x_c,$ where delocalized bosonic holons will experience a
Bose condensation at low temperatures$.$

\section{Experimental Implications}

\subsection{Thermopower}

The spontaneous translational symmetry breaking in a lightly doped Mott
insulator has very unique experimental consequences. In such a system, since
each doped hole can be self-trapped anywhere in space, due to the
translational symmetry of the original Hamiltonian, there will be a {\em %
large} and{\em \ doping-dependent entropy} associated with the energetically
degenerate configurations of holes in real-space distribution. Such an
anomalous entropy associated with the charge carriers can be directly probed
in a thermopower measurement.

It has been previously known \cite{heikes,chaikin} that for charge carriers
in a narrow band, when the temperature is raised to exceed the bandwidth,
the thermopower will be saturated to a $T$-independent constant, which is
entirely determined by the entropy change per added carrier. Namely, the
thermopower $S_e$ at high-$T$ will reduce to 
\begin{eqnarray}
S_e &\rightarrow &\frac \mu {eT}  \nonumber \\
&=&-\frac 1e\left( \frac{\partial S}{\partial N_e}\right) _E  \label{se}
\end{eqnarray}
where $\mu $ denotes the chemical potential, $S$ the entropy, and $N_e$ the
total electron number. The subscript $E$ here means that the partial
derivative is under a constant energy.

The self-localization of doped holes means a vanishing bandwidth of the
charge carriers and a spontaneous translational symmetry breaking with a lot
of degenerate real-space configurations in the distribution of holes. The
formula (\ref{se}) can be directly applied in such a case based on a general
thermodynamic consideration.

Putting $N_h$ hole on $N$ lattice sites with no-double-occupancy would give
rise to an entropy $S=$ $k_B\ln N!/N_h!\left( N-N_h\right) !,$ which then
leads to the Heikes formula \cite{heikes,chaikin} $S_e^H=\frac{k_B}e\ln 
\frac{(1-\delta )}\delta ,$ by noting $N_e=N-N_h$ in Eq.(\ref{se}) and using
the Sterlings approximation at $N\rightarrow \infty $ with $N_h=N\delta $.
But since each doped hole is actually a dipolar composite composed of a
meron and an antimeron located at two poles, which do not coincide with each
other as wall as with other hole dipoles, the total entropy should be
reduced by this fact. One can determine the entropy by assuming that two
poles of each hole dipole are loosely bound such that it becomes a problem
with a total number of $2N_h$ poles$,$ instead of $N_h,$ being put on $N$
lattice sites without double occupancy. Correspondingly one obtains the
following modified Heikes formula 
\begin{equation}
S_e^{mH}=\frac{k_B}e\ln \frac{(1-2\delta )}{2\delta }.  \label{se2}
\end{equation}
This formula has no other fitting parameters and is a universal function of
the doping concentration $\delta $.

\begin{figure}[tbp]
\begin{center}
\includegraphics{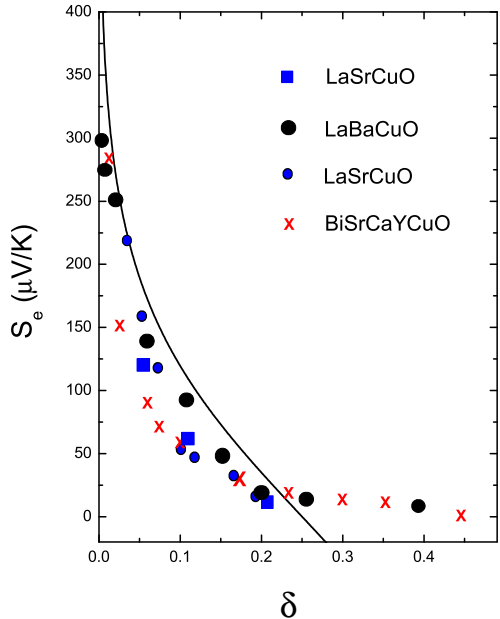}
\end{center}
\caption{The thermopower determined by Eq.(\ref{se2}) as a function of doping
concentration (solid curve). Experimental data are from Refs. \cite{coo,man,wang} (see text). }
\label{fig3}
\end{figure}

The Heikes-type formula (\ref{se2}) is plotted in Fig. 3, as the solid
curve, together with the experimental data obtained at room temperatures in the 
{\rm Sr} and {\rm Ba }doped {\rm La}$_2${\rm CuO}$_{4-y}$ compounds \cite{coo} 
(full square and bigger circle), {\rm %
Bi}$_2{\rm Sr}_2{\rm Ca}_{1-x}{\rm Y}_x{\rm Cu}_2{\rm O}_{8+y}$
system \cite{man} (cross), and from the recent
measurement by Wang and Ong \cite{wang} in {\rm La}$_2${\rm Sr }$_x${\rm CuO}$_{4-y}$
(small full circle). Fig. 3 shows that Eq.(\ref{se2}) agrees
qualitatively and quantitatively well with the experimental measurements in
the insulating regime without any fitting parameters, where the experimental
thermopower is sharply reduced from $300$ $\mu {\rm V/K}$ near half-filling
to around $0$ in the metallic regime.  
The agreement of the theory and
experiment quickly deviates in the metallic regime, where the experimental
thermopower remains within a narrow range of $\sim \pm 10$ $\mu {\rm V/K}$
in the optimal and overdoped regimes of the metallic phase and with a much
prominent temperature dependence. We also caution that in Fig. 3, the doping concentrations 
in the data for {\rm Bi}$_2{\rm Sr}_2{\rm Ca}_{1-x}{\rm Y}_x{\rm Cu}_2{\rm O}_{8+y}$ 
\cite{man} were indirectly determined by the method involving the Hall effect which 
may be not as reliable as the hole density obtained in {\rm LSCO} and {\rm LBCO} compounds.  

Note that some modified Heikes formulae have been used \cite{coo,man}
phenomenologically to fit the magnitude and doping-dependence of the
experimental data in the same low doping regime. However, it has long been a
puzzling question why the hole bandwidth should be shrunk to an order of
magnitude smaller than the temperature scale $\sim 100$ {\rm K }in order to
explain the experiment. The self-localization of doped holes in the present
theory, on the other hand, naturally explains this. 

An another important experimental fact is that $S_e$ has been generally
found \cite{ong,coo,man} to decrease continuously to zero as $T$ is reduced
below $100$ {\rm K} . Such a phenomenon can be easily understood in our
theory as follows. Since the holes are self-trapped in space, any
impurities, no matter how weak, can easily {\em pin down} them in space at
sufficiently low temperatures, truly breaking the translational invariance,
and therefore causing the diminishing of the degeneracy (and thus the
entropy). The thermopower should then quickly deviate its high-temperature
saturation value and vanish as $T\rightarrow 0$.{\rm \ }In this regime, the
Mott VRH will dominate the charge transport, as to be discussed below.

\subsection{Resistivity}

\subsubsection{Mott variable range hopping at low temperatures}

Experimentally, the cuprate superconductors have universally exhibited the
localization of charge carriers at low temperatures, in the low-doping
regime of $\delta <0.05$. The resistivity can be well fit \cite
{preyer,chen,kastner,ando} by the following Mott VRH formula 
\begin{equation}
\rho _M\sim e^{\left( \frac{T_0}T\right) ^{1/\gamma }}
\end{equation}
with $\gamma $ $\sim 3-4$ and $T_0\sim 10^6$ {\rm K} at $T\rightarrow 0,$
usually applicable to a doped semiconductor$.$ This implies a strong
localization of the doped holes in this regime. But lightly doped cuprates
by no means resemble a doped semiconductor. The strong Coulomb interaction
makes it a doped Mott insulator, in which the doped holes interact {\em %
strongly} with the spin background. As the result, they can be self-trapped
at low doping even without any disorder as described before.

So the self-localization of the doped holes in the lightly doped Mott
insulator will provide an intrinsic mechanism to explain the localization
phenomenon generally observed in the cuprates. Disorder or impurities, on the
other hand, should only play a minor role in such a system. As noted above,
in the presence of disorder, the spontaneous translational symmetry breaking
of the lightly doped Mott insulator (with a lot of degeneracies) can easily
become truly translational breaking, as the self-localized holes, without
the penalty from the kinetic energy, are energetically in favor of staying
near the impurities. Therefore, the low-doping phase can also be regarded as
a strong Anderson localization system at low temperatures, even though the
presence of disorder or impurities may not be really strong. In other words, 
{\em the impurity effect will get ``amplified'' by the Mott physics at low
doping.}

Recall that a holon has its own bare hopping term, governed by $H_h$\ in Eq.(%
\ref{hh}), which in the continuum limit reduces to 
\begin{equation}
H_h\simeq \int d^2{\bf r}\frac{(-i\nabla +{\bf A}^s)^2}{2m_h},  \label{hhc}
\end{equation}
with an effective mass $m_h=\frac 1{2t_ha^2}$\ and ${\bf A}^s$\ as the
continuum version of the gauge field $A_{ij}^s$.{\bf \ }So the holon is
expected to hop around based on $H_h$\ and is bound to the induced antimeron
by the attractive potential (\ref{vdipole}). At low temperatures, ${\bf A}^s$%
\ may be neglected as spinons are in RVB pairing \cite{string1}.{\bf \ }Then
the Schr\"{o}dinger equation for a hole-dipole can be written down by 
\begin{equation}
-\frac{{1}}{2m_h}{\bf \nabla }^2\psi +V\psi =E^h\psi .
\end{equation}
Define $\psi ({\bf r})\equiv \psi (r,\phi )$.\ To compute the radial
component of the wave function we note that asymptotically at large $r$\ the
Schr\"{o}dinger equation reduces to a form whose (radial) solutions can be
expressed as\ 
\begin{equation}
\psi (r,\phi )~\sim e^{-\sqrt{\ln \frac r{a_0}}\frac r{a_0}}  \label{psi}
\end{equation}
where $a_0=\frac 1{\sqrt{2{m}_hq^2}}=\sqrt{\frac{t_h}{\pi \tilde{J}_s}}a$.

Then one may estimate the transition probability $\Gamma _{ij}$ of the holon
between any two adjacent antimerons (``impurities''), located at $i$ and $j,$
based on Eq.(\ref{psi}). It is given by 
\begin{equation}
\Gamma _{ij}{\bf \sim }\exp ({\bf -}\frac{2r_{ij}\sqrt{\ln \frac{r_{ij}}{a_0}%
}}{a_0}{\bf -}\frac{\epsilon _{ij}}T)\text{ }  \label{gamma}
\end{equation}
where $r_{ij}$ is the distance and $\epsilon _{ij}$ is the on-site energy
difference between two ``impurity'' sites. Except for the factor $\sqrt{\ln 
\frac{r_{ij}}{a_0}},$ this formula is essentially the same as in the
original Mott theory. As the temperature is lowered, the motion between
neighboring sites becomes more difficult due to the lack of appropriate
energy differences. Consequently, it is more likely for the carriers to hop
to a more distant site if this means that the energy difference is less. It
is known \cite{mott,mott1,goodingSG} as the Mott VRH. Except for a
logarithmic correction, the resistivity for the 2D Mott VRH can be
determined according to Eq.(\ref{gamma}) by the following expression 
\begin{equation}
\rho (T)\sim \,e^{\left( \frac{T_0}T\right) ^{1/3}\sqrt{\frac 13\ln \frac{T_0%
}T}}  \label{sigma}
\end{equation}
where $T_0$ is a characteristic temperature given by $T_0=\frac{13.8}{%
D_0a_0^2}.$ Here $D_0$ is the energy density of the impurity states, which
is assumed to be constant in the VRH regime.

It has been well known that once the anisotropic 3D is considered, the
exponent $\gamma $ in the Mott VRH conductivity generally will be changed
from $3$ to $4$, with a modified $T_0$\cite{sh}$.$ Furthermore, the
interacting effect between the holons, given by $V_{12}=-\pi \tilde{J}_s\ln 
\frac{\mid {\bf r}_1{\bf -r}_2{\bf \mid }}a$, has been ignored here, which
can also modify the exponent in the VRH theory.

\subsubsection{Crossover to deconfinement at high-temperatures}

The ``deconfining'' temperature $T_{{\rm de}}$ will represent a
characteristic temperature beyond which holons are deconfined from the
antimerons. Once the holons are unbound from their antimeron partners and
move freely at $T>T_{{\rm de}}$, their transport will be solely governed by
the hopping term of the phase-string model (\ref{hhc}). Here the interaction
between holons and antimerons becomes irrelevant as $V\rightarrow 0$ in the
above RG analysis. Instead we must consider the contribution from the gauge
fluctuations of ${\bf A}^s$ in Eq.(\ref{hhc})$,$ which will play an
essential role for scattering at high temperatures.

Note that ${\bf A}^s$ satisfies the following condition: 
\begin{equation}
\oint_Cd{\bf r\cdot A}^s=\pm \pi \sum_{l\in \Sigma _C}\left( n_{l\uparrow
}^b-n_{l\downarrow }^b\right) \equiv \Phi _C.  \label{fluxs}
\end{equation}
We can estimate the strength of the fluctuations of ${\bf A}^s$\ by defining 
\begin{equation}
\varpi =\sqrt{\langle \Delta \Phi _{\Box }^2\rangle },
\end{equation}
where $\Phi _{\Box }$\ denotes the flux per plaquette: 
\begin{equation}
\Phi _{\Box }=\pm 2\pi \frac 14\sum_{\Box }S_l^z
\end{equation}
according to Eq.(\ref{fluxs}). At very high-temperature limit, one may
neglect the NN spin-spin correlations such that 
\begin{eqnarray}
\varpi &\sim &2\pi \sqrt{<(S_l^z)^2>}  \nonumber \\
&=&\pi \sqrt{1-\delta }\sim \pi ,  \label{f}
\end{eqnarray}
which implies very strong flux fluctuations per plaquette in the high-$T$
limit.

As a matter of fact, if the fluctuations of ${\bf A}^s$ is treated in the
quasistatic limit with an annealed average over static flux distributions
in Eq.(\ref{hhc}), the transport properties are the same as those
studied in Ref. \cite{lee}. In particular, the scattering rate has been
found \cite{lee} 
\[
\frac 1\tau \simeq 2k_{{\rm B}}T, 
\]
if $\varpi >\pi /2$, which is satisfied in our case according to Eq.(\ref{f}%
) in the high-$T$ limit. Corresponding, the resistivity is 
\begin{equation}
\rho \sim T\text{ \ \ \ \ at }T\gg 1.  \label{rho}
\end{equation}
Namely, the charge transport in high-$T$ ``deconfining phase'' will
generally follow a strange-metal behavior due to the scattering between the
holons and gauge field ${\bf A}^s$ in Eq.(\ref{hhc}).

In the crossover from the low-$T$ VRH behavior (\ref{sigma})
to the high-$T$ linear temperature behavior (\ref{rho}), one expects to see
a minimal resistivity $\rho _{{\rm \min }}$. We point out that generally
the ``deconfining temperature'' $T_{{\rm de}}$ does not necessarily coincide
with the characteristic temperature of $\rho _{{\rm \min }}$. The latter may
occur at a lower temperature as the fluctuations in ${\bf A}^s$ can already
become important when the N\'{e}el temperature $T_N$ is approached from
below. In the above discussion of the VRH resistivity, such a scattering
effect has been neglected at low-$T$, which should lead to the enhancement of the
resistivity once it becomes important. Another possibility is that the holon
induced antimerons may start to move with the increase of temperature,
resembling the ``flux-flow'' in a superconducting phase, which also may lead
to a qualitative change of the resistivity. These possibilities in the
intermediate temperature regime are beyond the scope of the present work.

\subsection{Dielectric constant}

Another interesting prediction of the self-localization of doped holes is
the existence of a {\em large }and {\em doping-dependent dielectric constant}
as each hole is a dipolar object of a bound state of a holon and a localized
antimeron. One thus expects that the dielectric constant increases linearly
with doping initially and finally diverges as the deconfining point is
approached with increasing doping concentration.

\begin{figure}[tbp]
\begin{center}
\includegraphics{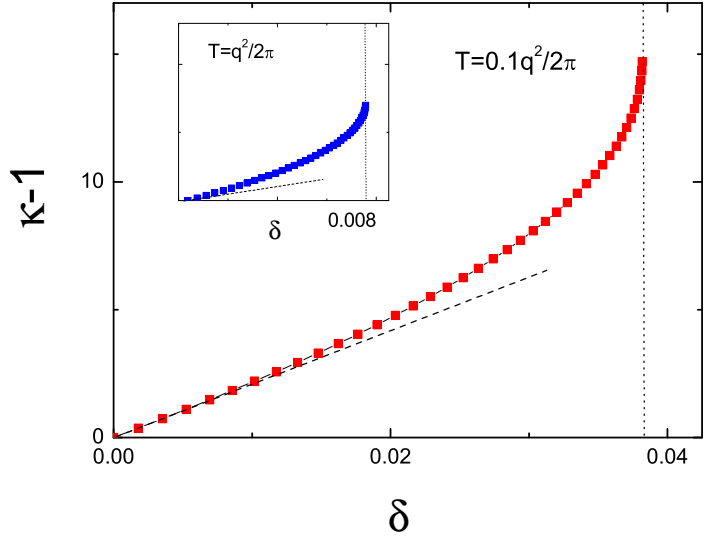}
\end{center}
\caption{The dielectric constant $\kappa$ as a function of doping calculated at different 
temperatures. $\kappa =1$ at half-filling, and the dashed lines indicate the linear 
dependence of the doping concentration. In the main panel, $\kappa$ diverges at 
$\delta_{\rm{de}}=0.038$ and in the inset $\delta_{\rm{de}}=0.008$.}
\label{fig4}
\end{figure}

The dielectric constant defined in the RG analysis in Sec. II C can be
written by $\kappa (l)=X(l)\beta q^2/2\pi .$ We can determine $\kappa
=\kappa (l=\infty )=X(l=\infty )\beta q^2/2\pi $ by Eq. (\ref{rc3})
numerically (taking $\kappa (l=0)=1$). Fig. 4 and the inset show the results
for $\kappa -1$ at $T=0.1\frac{q^2}{2\pi }$ and $T=\frac{q^2}{2\pi },$
respectively. At half-filling, $\kappa =1$ as no contribution from the
hole-dipoles. We see that with the increase of the hole concentration, the
dielectric constant grows linearly with $\delta $ at first, then deviates
the linearity shown by the dashed line, indicating the increase of the
dipole size. Eventually it diverges at a critical doping concentration $%
\delta _{{\rm de}}=\delta _{{\rm de}}(T),$ beyond which the dipoles will
collapse and free holons will be released.

The general trend of the calculated in-plane dielectric constant, shown in 
Fig. 4, qualitatively agrees \cite
{remark2} with the experimental
measurements \cite{chen,kastner} in the low-doping cuprates. Indeed, for the
lightly-doped cuprates, a large, doping-dependent dielectric constant has
been observed in the $ab$-plane, which increases with the hole concentration 
$\delta $, initially linearly then becoming divergent at some higher
concentrations \cite{chen,kastner}. In contrast, the out-plane (c-axis)
dielectric constant shows no essential change as a function of $\delta $.

In a doped-semiconductor picture, a dielectric constant contributed by the
doped holes can only be significant when the holes are bound to impurities,
which is in the localized regime at low temperatures. But such a dielectric
constant should be usually anisotropic 3D-like rather than pure 2D-like as
the experiment revealed. Especially it is difficult to explain why the
dielectric constant should diverge in the $ab$-plane while remains constant
in the c-axis. Furthermore, if the majority of the holes remains bound to
impurities, it is hard to reconcile with the large{\em \ saturated}
thermopower observed at $T\gtrsim 100$ {\rm K}.

In contrast, the dipolar structure of the doped holes in the present
framework can naturally lead to a {\em large} dielectric constant {\em in
the }$ ab${\em -plane}, no matter whether the hole-dipoles are pinned
down by impurities or not, as long as the hole dipole composites remain
stable. The hole dipoles are presumably de-pinned from impurities at $%
T\gtrsim 100$ {\rm K }in our theory, since a large thermopower has been seen
in experiment. With further increasing temperature, the dipolar structure
will eventually collapse and the holon will become {\em deconfined} from its
antimeron partner. Consequently the large dielectric constant and
thermopower should be both quickly reduced above $T_{{\rm de}},$ where the
resistivity also starts to behave like a strange-metal as $T$ becomes
sufficiently high.

\section{Conclusions}

In this paper, we studied the motion of doped holes in a spin ordered
background at low doping. Based on the phase-string model, we demonstrated
that the holes will get self-localized in space, leading to a spontaneous
translational symmetry breaking without the presence of disorder or
impurities.

This novel property is an important consequence of the Mott insulator at low
doping, described by the phase-string model as the low-energy effective
description of the $t-J$ Hamiltonian. The doping effect and the interplay
between charge and spin degrees of freedom are characterized by a unique
gauge structure with a mutual duality. At low doping, the spinon
condensation forces a ``confinement'' on the holons, making the latter
self-localized and resulting an insulator with an AFLRO or spin glass. This
is in contrast to the higher-doping phase, where the holon condensation
forces a ``confinement'' on the spinon part, resulting in a superconducting
phase coherence \cite{zhou}.

We found strong experimental implications based on the self-localization of
holes. Large and doping-dependent thermopower can naturally explain the
experimental data which had been very hard to understand by conventional
theories. A large and doping-dependent in-plane dielectric constant
indicates a composite structure of the holes and provides a unique
explanation of the experimental observations, which otherwise are very
difficult to comprehend. Furthermore, the low-$T$ VRH resistivity observed
experimentally was interpreted as the direct consequence of the
self-localization with disorders playing a minor role, which explains why
the critical doping of the delocalization in the cuprates is universally
around $\delta _c\sim 0.05$ at $T=0,$ not very sensitive to the density of
disorders in the samples. The phase-string model also naturally shows how
the resistivity evolves into a strange-metallic linear-$T$ behavior at
sufficiently high temperatures above the delocalization temperature. Most
importantly, we wish to emphasize that all these peculiar experimental
properties were shown to be consistently explained within a single
theoretical framework.

It should be noted that many results in this paper are only correct for
a homogeneous phase. There may exist another possibility, namely, the stripe
instability \cite{kou1} in the phase string model, which can result in an
inhomogeneous phase. Since holes are self-localized, their kinetic energies
are suppressed such that the potential energy will become predominant. The
dipole-dipole interaction might cause stripe instability at low
temperatures, with hole-dipoles collapsing into a one-dimensional line-up
(stripe) \cite{kou1}. The pinning effect of disorders may stabile the
homogeneous phase at low-$T,$ so does the long-range Coulomb interaction.
But it would be very interesting to incorporate the inhomogeneous tendency in
various dynamic properties at low doping in future investigations.

\vskip 1cm

\acknowledgements
We would like to thank Y. Ando, N.P. Ong, and M.L. Ge for helpful
discussions, and particularly Y. Wang for kindly sending us their unpublished 
experimental data and useful comments. SPK acknowledges support from NSFC and 
ZYW acknowledges support from NSFC 
and MOE of China.


\begin{references}
\bibitem{remark1}  One may equivalently get a degenerate AFLOR state with
the staggered magnetic moment pointing at $z$-axis. See Ref.\cite{aa}.
\bibitem{remark2}  We note that the dielectric constant $\kappa $ defined in
the present case may also be expressed by $\kappa =1+2\pi \rho _hp$ ($\rho
_h=\delta /a^2),$ in which the polarizability $p$ is determined as follows.
Applying an electric field{\bf \ }${\cal E}$ at an angle $\theta $ to the
line of $r$ connecting the holon and the antimeron, then it can be shown 
\cite{BKT} $p=q\frac{\partial <r\cos \theta >}{\partial {\cal E}}\mid _{%
{\cal E}=0}$ with a ``charge'' $q$. On the other hand, the dielectric
constant  $\epsilon$ determined in experiment should be related to a polarizability in
which it involves a true charge $e$ and in general one expects $\epsilon
-1\propto \kappa -1.$
\end{references}
\end{document}